\documentclass[a4paper]{jpconf}
\usepackage{graphicx}
\usepackage{amsmath}

\usepackage{epsfig}

\usepackage{graphicx}
\usepackage{dcolumn}
\usepackage{bm}
\usepackage{epstopdf}

\newcommand{\fgb}{f_{\! G}^b}

\renewcommand{\sl}{{\it sl}}
\newcommand{\wl}{{\it wl}}

\newcommand{\rms}{r_{\it ms}}

\newcommand{\eVct}{eV/$c^2$}
\renewcommand{\eVct}{{\rm eV}}

\renewcommand{\ms}{{\it ms}}

\newcommand{\gbar}{{ g}}
\newcommand{\myskip}[1]{}

\newcommand{\BEQ}{\begin{eqnarray}}
\newcommand{\EEQ}{\end{eqnarray}}
\newcommand{\BEA}{\begin{eqnarray}}
\newcommand{\EEA}{\end{eqnarray}}

\newcommand{\Sigmab}{\overline\Sigma}

\newcommand{\cm}{{\rm cm}}
\newcommand{\gr}{{\rm gr}}
\newcommand{\km}{{\rm km}}

\newcommand{\s}{{\rm s}}

\newcommand{\kpc}{{\rm kpc}}

\newcommand{\eV}{{\rm eV}}

\newcommand{\cg}{{\it cg}}
\newcommand{\co}{{\it co}}

\newcommand{\Gal}{{\it G}}

\newcommand{\Om}{\Omega}

\newcommand{\LCDM}{{$\Lambda$CDM}}

\newcommand{\figSig}{1}
\newcommand{\figgt}{2}

\begin{document}

\title[Neutrino mass and signature]{Dirac neutrino mass from a neutrino dark matter model for the galaxy cluster Abell 1689}

\author{Theodorus Maria Nieuwenhuizen}

\address{$^1$Institute for Theoretical Physics,  University of Amsterdam, Science Park 904, 1090 GL  Amsterdam, The Netherlands \\
$^2$International Institute of Physics, Federal University of Rio Grande do Norte, Natal, Brazil}


\newcommand{\masscorr}[1]{}

\begin{abstract}
The dark matter in the galaxy cluster Abell 1689 is modelled as an isothermal sphere of neutrinos. 
New data on the $2d$ mass density allow an accurate description of its core and halo.
The model has no  ``missing baryon problem'' and beyond $2.1$ Mpc the baryons have the cosmic mass abundance. 
Combination of cluster data with the cosmic dark matter fraction -- here supposed to stem from the neutrinos --
leads to a solution of the dark matter riddle by left and right handed neutrinos with mass 
\masscorr{$(1.861\pm0.016)h_{70}^{-2}\eV/c^2$ {\bf NO, NO, IT IS} }
$1.847\pm0.016 \, \eV/c^2$.
The thus far observed absence of neutrinoless double beta decay points to (quasi-) Dirac neutrinos: uncharged electrons with different flavour 
and mass eigenbasis, as for quarks. Though the cosmic microwave background spectrum is matched up to some 10\% accuracy only,
the case is not ruled out because the plasma phase of the early Universe may be turbulent.
\end{abstract}


\hfill{\it Nothing matters more than dark matter\footnote{And nothing energises more than dark energy}}

\section{Introduction}

The status of dark matter (DM) is paradoxical.
While the cosmic microwave background (CMB) \cite{hinshaw2013nine,planck2015planck} and baryon acoustic oscillations 
\cite{anderson2012clustering} bring strong support for a WIMP (weakly interacting massive particle), 
the axion or perhaps a sterile neutrino,  three decades  of (in)direct searches have consistently left the researchers empty handed.
Recent experiments include LUX \cite{akerib2014first}, an underground Xe experiment, the Large Hadron Collider (LHC) in its 2012-2014
 run at 7-8 GeV \cite{baer2014susy}, the Alpha Magnetic Spectrometer at the International Space Station measuring electrons, positrons and (anti)protons
\cite{accardo2014high, giesen2015ams}, and the Fermi X-ray telescope \cite{ackermann2011constraining,fermi2015searching,wood2015dark}.
None of them yielded a decisive clue. Hopes are now set on the 13 TeV run at the LHC, till now again without dark matter hints,
the next one of Lux and the first of Xenon1T.

In parallel with this, the 2-12 GeV gamma-ray excess radiation in the Galactic center is likely not due to dark matter but of astrophysical origin \cite{lee2015evidence,bartels2015strong},
while the possibility of a 7 keV warm dark matter particle (sterile neutrino) related to a 3.5 keV X-ray line observed first in Andromeda and the Perseus 
cluster \cite{boyarsky2014unidentified}  is fading now that it is not observed in Draco \cite{jeltema2015deep,ruchayskiy2015searching}.

But are we perhaps barking up the wrong tree?
Various tensions with  $\Lambda$ cold dark matter (\LCDM) cosmology are known, let's mention a few.
Planck confirms the existence of anomalies, in particular the cold spot \cite{rassat2014planck}, which may be aligned with a supervoid \cite{szapudi2015detection};
on a scale of nearly a giga parsec, 19 quasars have been observed with spins likely aligned with their hosts large-scale structures \cite{hutsemekers2014alignment};
 a ring of nine gamma ray bursts spreads more than 5 billion light years across, exceeding the 
 theoretical limit of 1.2 billion light years for the largest possible structures expected from the cosmological principle \cite{balazs2015giant};
for the ``cosmic train wreck" galaxy cluster Abell 520, which has a central $3-4\,10^{13}M_\odot$ 
mass clump with the enormous mass-to-light ratio $800M_\odot/L_{R\odot}$, it is difficult to explain why weakly interacting DM would 
have separated itself from the galaxies \cite{jee2012study,clowe2012dark,jee2014hubble};
in the cluster Abell 3827 an offset between baryonic and dark mass occurs \cite{massey2015behaviour}, 
which is an order of magnitude larger than \LCDM \  can explain \cite{schaller2015offsets}.
Global issues exist as well: the predicted transition for the most massive galaxies to transform from their initial halo assembly at redshifts $z=8-4$ 
to the later baryonic evolution  seen in star-forming galaxies and quasars is not observed \cite{steinhardt2015impossibly};
the galaxy power spectrum deduced from SDSS-III observations fits well to the stretched exponential $\exp[-(k/k_b)^{1/2}]$ 
predicted by turbulence  \cite{bershadskii2015deterministic}.

These and further issues \cite{nieuwenhuizen2009gravitational,nieuwenhuizen2010micro,nieuwenhuizen2011explanation}
make us reflect on the whole DM riddle. Of course, the most natural source for DM is neutrinos. 
In \LCDM \  they cannot make up the DM due to free streaming: 
structure formation is hindered when the total mass in the neutrino families exceeds  a bound that is maximally 
0.72 eV/$c^2$ \cite{planck2015planck}. This corresponds to a cosmic mass fraction $\Omega_\nu< 1.6\%$, far
below the cosmic dark matter fraction $\Om_c=24.2\pm0.2\%$ 
\footnote{Unless indicated otherwise, we adopt the cosmology 
$H_0=h_{70}70$ km/s Mpc, $h_{70}=1$, $\Omega_m$ = $0.30$ and $\Omega_\Lambda=0.70$.
One arc minute corresponds  at the redshift $\bar z=0.183$ of A1689  to  $1'=184.5$ kpc.}.
However, there is an argument against this {\it linear} structure formation: 
{\it the Jeans instability is absolute} beyond linear perturbations, since any domain of ``Jeans-stable'' plane waves contains pairs of plane waves which produce 
at the bilinear level a beat with wavelength inside the Jeans-unstable regime \cite{Nieuwenhuizen2016Jeans}.
Hence the plasma may be (weakly) turbulent, produce nonlinear structures like galaxy clusters \cite{Gibson1996Turbulence},
explain many perplexing observations \cite{nieuwenhuizen2009gravitational,nieuwenhuizen2010micro}
and leave signatures of turbulence in the galaxy distribution \cite{bershadskii2015deterministic}.
Indeed, the quark-gluon plasma has a low viscosity, which leads to a very large Reynolds number Re $\sim$ $10^{19}$ for a patch of acoustic 
horizon size  \cite{Nieuwenhuizen2014Natal,Nieuwenhuizen2015Prague}. 
Since this horizon is larger than the causal horizon,  turbulence can not develop --  at {\it this} scale.
But  {\it turbulence can develop on subhorizon scale} -- and hence it should if there is enough time. 
The plasma keeps large Reynolds numbers till $\sim 100$ at the recombination, so it is expected to remain turbulent untill then \cite{Nieuwenhuizen2014Natal,Nieuwenhuizen2015Prague}.
Hence, lacking further study,  the option that neutrinos make  up the dark matter, is not ruled out by the neutrino free streaming in linear approaches.
We shall not dwell further into this, but rather test whether the case is indeed worth the effort.

Neutrino oscillations yield small $m^2$ differences, so if they are heavy,
they have nearly the same mass. The upper limit of the electron antineutrino mass is 
$2 .0$ eV/$c^2$  \cite{Agashe:2014kda}, and for the 3 families ($\nu_e$, $\bar\nu_e$;
$\nu_\mu$, $\bar\nu_\mu$; $\nu_\tau$, $\bar\nu_\tau$) the thermal densities
add up to  $\Omega_\nu=\rho_{\nu}/ \rho_c\le 0.13$, less than $\Om_c$.
Hence if $\nu$'s are the source of DM, also right-handed ones (sterile $\nu$'s, not involved in weak processes) 
must exist. Reactor experiments favor them too \cite{giunti2007fundamentals,schwetz2011global,kopp2011there}.

Cosmology offers further evidence. While the CMB has achieved  the status of ``precision cosmology'', gravitational lensing can reach this status too.
Indeed, the quite relaxed galaxy cluster Abell 1689 offers a stringent test. 
There are  well constrained data for the $2d$ mass density $\Sigma(r)$, primarily from strong lensing (SL),
that is to say, from background galaxies lensed into several ($\le5$) pieces of an arc, up to some 200 kpc off the cluster centre.
Further out, weak lensing (WL) occurs, that is, randomly oriented background galaxies get a systematic tangential shear $g_t(r)$ towards
a circle around the cluster centre \cite{bartelmann2001weak}.
The effect can be deduced by averaging over the galaxies within a suitably defined field. 
The cluster galaxies can of course be observed easily. Finally, X-ray observations 
provide the electron density profile $n_e(r)$ and thus the mass density of the X-ray gas.  These inputs
provide a unique basis for an accurate reconstruction of the mass profile. 

The setup of this paper is as follows. First we shortly discuss A1689 and the data sets which are analyzed.
Next we introduce model for the baryons in the Brightest Cluster Galaxy (the central one), the only relevant galaxy for our purpose,
and we fit the gas data to a S\'ersic profile. This is followed by a fit to the NFW profile of dark matter. Next, a model for neutrino dark
matter is introduced and developed and a fit is presented. The results are combined with results from Planck and neutrinoless double beta decay.
The paper closes with a summary.

\section{The galaxy cluster Abell 1689}

Abell 1689 is a rather large, well relaxed galaxy cluster that has long been studied.
An initiating WL study of A1689 was performed in \cite{tyson1995measurement}. Ref. \cite{limousin2007combining}
reports a SL analysis yielding $\Sigma$ for radii up to 270 kpc.
Ref. \cite{umetsu2008combining} employs an entropy-regularized maximum likelihood analysis of the lens magnification and the distortion 
 of red background galaxies detected with Subaru.
Ref. \cite{morandi2011reconstructing} unveils the $3d$ structure of galaxy clusters such as A1689 and resolves the discrepancy between X-ray and lensing masses. 
Ref. \cite{umetsu2015three} fits the NFW profile of cold dark matter \cite{navarro1997universal} with the concentration parameter $c_{200}$ = $8-9$ \cite{umetsu2015three}.
However, it is known that modelling only the central data $r<300$ kpc leads to smaller values, $4-5$.
An NFW fit of the combined data leads to $c_{200}=4.5$ \cite{nieuwenhuizen2013observations}.
Nowadays the focus is on the trixiality of A1689 \cite{morandi2011reconstructing,sereno2011weak,umetsu2015three}.
 However,  the spherical approximation will serve our goals,
 since most of the matter is dark and more spherical than the gas.
   
Most authors focus on the NFW profile,  so that no information about the mass of the DM particle is obtained.
Still, some groups consider isothermal fermions.  Cowsik \& McClelland  \cite{cowsik1973gravity} 
model the DM of the Coma cluster
as an isothermal sphere of neutrinos,  which leads to a mass\footnote{In the text we employ the short hand eV  for eV/$c^2$.} of $\simeq 2$ eV. 
Treumann et al.  study  2 eV \ thermal neutrinos next to thermal CDM and X-ray gas for clusters like Coma \cite{treumann2000neutrino}.
We apply an isothermal fermion model for a single type of dark matter, for the galaxies and the X-ray gas;
a fit to lensing data of the Abell 1689 cluster works well and yields as best case the neutrino with mass  $\sim1.5$ eV  \cite{nieuwenhuizen2009non}. 
This approach has been followed up whenever new data became available.   
In \cite{nieuwenhuizen2011prediction} we apply the model to SL data by \cite{coe2010high} and 
X-ray data by \cite{morandi2010unveiling}, confirming the result.
In \cite{nieuwenhuizen2013observations} we adopt the galactic matter profile of \cite{limousin2005constraining},
to  find that neutrinos perform better ($\chi^2/\nu\sim0.7$) 
 than the NFW profile ($\chi^2/\nu\sim2.2$).

\subsection{Data sets to be employed.}  
Recently, Umetsu et al. perform a WL analysis combined with some magnification properties \cite{umetsu2015three}, 
which provides important  information as we shall discuss below.
The authors provide 14 data points $\Sigma^i_\wl$ at $124<r_i<2970$ kpc and
the correlation matrix $C_{ij}^\wl$. The latter needs no regularisation, because each bin has $S/N>1$.
The methods for combining the WL shear and magnification effects to derive the radial mass density profiles 
are explained in \cite{sereno2011weak,umetsu2011cluster}.
Between 124 and 270 kpc the data overlap with those of \cite{limousin2007combining}.

We take additional data for $\Sigma(r)$  from SL analysis by Limousin et al \cite{limousin2007combining}.
For description of the analysis we also refer to \cite{nieuwenhuizen2013observations}.
There result $12$ data points $(r_i,\Sigma_\sl^i)$ for $3<r_i<270$ kpc and their covariance matrix $C_{ij}^\sl$,
which has eigenvalues between $0.03\Sigma_c^2$ and $1.4\,10^{-8}\Sigma_c^2$.
To account for information due to (implicitly assumed) priors and for further scatter, we add a constant $\sigma_\sl^2$ to the diagonal elements,
so that  the role  of the small eigenvalues is suppressed. This leads to the modified covariance matrix  
$\overline{ C}_\sl =\sigma_\sl^2I+C_\sl$. A model for $\Sigma(r)$ then involves
$\chi^2_\sl(\Sigma)=\sum_{i,j=1}^{12}[\Sigma(r_i)-\Sigma_{\sl}^{i}] (\overline{C}_\sl^{\,-1})_{ij}[\Sigma(r_j)-\Sigma_\sl^j]$.
 We choose $\sigma_\sl$ on an empirical basis. In order not to erase the information
coded in $C_\sl$,  one needs $\sigma_\sl^2\ll 0.03\Sigma_c^2$.
In \cite{nieuwenhuizen2013observations} we take the average of the diagonal elements,  now we even allow half of it:
$\bar\sigma_\sl^2={\rm tr}\,C_\sl /24=(0.0375\Sigma_c)^2$.

We also employ the 13 data points for $g_t$  between 200 kpc and 3 Mpc from the WL study \cite{umetsu2015three}.
While based on better statistics, they overlap with the data of \cite{umetsu2008combining}. 

The data for $n_e$ we take from  Morandi et al \cite{morandi2010unveiling},  who analyze Chandra X-ray observations.
The density profile is recovered in a non-parametric way by rebinning the surface brightness into circular annuli 
via spherical deprojection \cite{morandi2007x}, see also the discussion in \cite{nieuwenhuizen2013observations}.

\section{ Modelling the baryons}

 In a galaxy cluster there are three components: Galaxies, X-ray gas and dark matter.
The galaxy mass density is dominated by the brightest cluster  galaxy (BCG, ``central galaxy'' $\cg$).
An adequate profile with mass $M_G$, core size $R_G^i$ and scale $R_G^o$
is \cite{limousin2005constraining}

\renewcommand{\co}{{\it Gi}}
\renewcommand{\cg}{{\it Go}}
\newcommand{\bcg}{{\it G}}

\BEQ
\rho_\Gal(r) = 
\frac{M_G ( R_G^{i}+R_G^{o})}{2\pi^2(r^2+R_G^{i\,2})(r^2+R_G^{o\,2})}.
\EEQ
If $R_G^i\ll R_G^o$ it is a cored, truncated isothermal sphere.

The mass density of the X-ray gas  follows \cite{morandi2010unveiling} as $\rho_g(r)=1.167\, m_Nn_e(r)$ 
for a typical $Z=0.3$ solar metallicity \cite{nieuwenhuizen2009non}.
The 56 data points for $n_e$ with the maximum of the upper and lower errors
fit  quite better to a  cored S\'ersic profile than to a $\beta$-profile. The S\'ersic profile has the form
  
\BEQ \label{nefit}
n_e(r)=n_e^0\exp\left[k_g-k_g\left(1+\frac{r^2}{R_g^{2}}\right)^{1/(2n_g)}\right].
\EEQ
The  best fit parameters are 

\BEQ \label{gasfit}
n_e^0=0.0670 \pm  0.0028\,\cm^{-3},\quad k_g=1.98\pm 0.25,  \quad  R_g= 21.6\pm2.7\, \kpc,\quad n_g=2.97 \pm 0.14. 
\EEQ
The value $\chi^2(n_e)/\nu=1.71$ with $\nu=52$ implies the marginally acceptable $q$-value 0.0010 for our spherical approximation.
A fit to a $\beta$ model yields $\chi^2(n_e)/\nu=6.3$, far from acceptable.

\begin{figure}
\label{figWLUBdataa}
\centerline{ \includegraphics[scale=1]{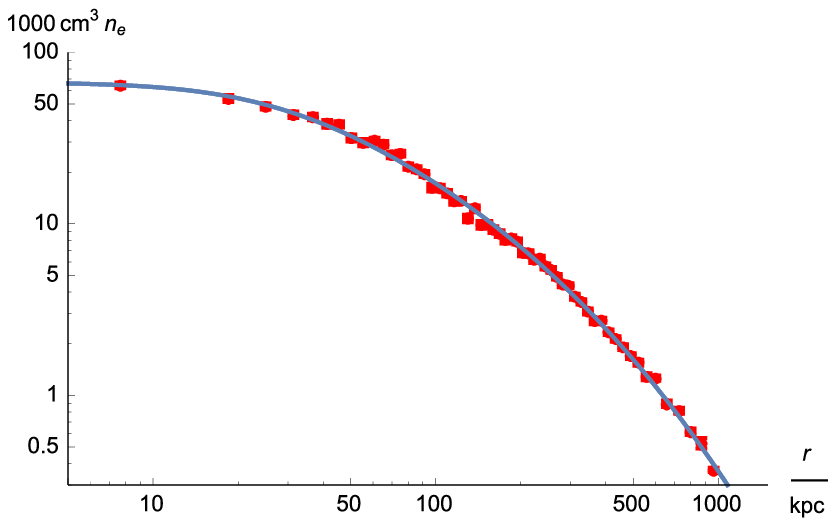}}
\vspace{-14mm}
\centerline{\includegraphics[scale=1]{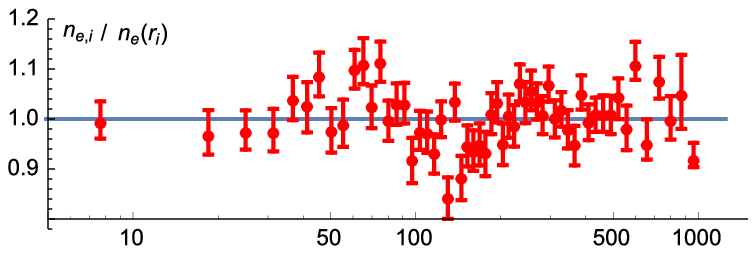}}
 \caption{Electron density as function of radius. Data from \cite{morandi2010unveiling}, full line: cored S\'ersic profile (\ref{nefit}). 
 Lower pane: relative values.}
 \end{figure}

\subsection{Towards data fits.}
SL data yield  the $2d$ mass density

\BEQ \label{Sigma-rho}
\Sigma(r)=\int_{-\infty}^\infty{\rm d} z\rho\left(\sqrt{r^2+z{}^2}\,\right).
\EEQ
Within a disc of radius $r$ it has the average 
$\Sigmab(r)=2r^{-2}\int_0^r{\rm d} r'\,r'\Sigma(r')$. 
In weak lensing one determines the transversal shear, which is related to $\Sigmab$ and $\Sigma$ as  

\BEQ\label{gt=}
g_t(r)=\frac{\Sigmab(r)-\Sigma(r)}{\Sigma_c-\Sigma(r)}.
\EEQ
The data of \cite{umetsu2015three} involve $\Sigma_c=0.682 \, h_{70}\, \gr \, \cm^{-2}$.

We fit our parameters by minimizing  $\chi^2(\Sigma,g_t,n_e)=\chi^2(\Sigma,g_t)+\chi^2(n_e)$, where
the lensing errors add up to

\BEQ \label{chisqtot=}
\chi^2(\Sigma,g_t)=\chi^2_\sl(\Sigma)+\chi^2_\wl(\Sigma)+\chi^2_\wl(g_t). 
\EEQ

\section{NFW fitting.}
Minimising only $\chi^2_\wl(\Sigma)+\chi^2_\wl(g_t)$ leads to $\nu=27-2$ and $\chi^2/\nu=1.25$, implying the good $q$-value $0.18$,
in support of a recent Bayesian analysis \cite{umetsu2015three}. 
However, the SL data put stress on this.
Ref. \cite{nieuwenhuizen2013observations} reports $[\chi^2_\sl(\Sigma)+\chi^2_\wl(g_t)]/\nu=2.2$ for its regulator $\sigma_\sl=\sqrt{2}\bar\sigma_\sl$ and $\nu=20$
so that $q=1.5\,10^{-3}$, marginally acceptable; this case now yields $\chi^2(\Sigma,g_t)/\nu= 2.21$ for $\nu=37$ with a bad $q$-value $3.3\,10^{-5}$.
Our present $\bar\sigma_\sl$  yields $\chi^2(\Sigma,g_t)/\nu=2.98$ with $\nu=37$ and $q=3.3\,10^{-9}$, very bad.
All by all, NFW fits are less satisfactory for the combined SL and WL data.

\section{Neutrino model.} 

For the DM we now consider  $g$ identical, nonrelativistic thermal  fermion modes, denoted by $\nu$, with mass $m_\nu$
and chemical potential $m_\nu\mu$, at temperature $T_\nu=m_\nu\sigma_\nu^2$. 
With potentional energy $m_\nu\varphi(r)$ due to all gravitating matter, its mass density reads 

\BEQ \label{nudens}
\!\!
\rho_{\nu}(r)\!=\!\!\int \! \!\frac{{\rm d}^3p}{(2\pi\hbar)^3}\frac{gm_\nu}{\exp\{[p^2/2m_\nu+m_\nu\varphi(r)-m_\nu\mu]/m_\nu\sigma_{\nu}^2 \}+1}.
\EEQ
Because positive energy particles will have escaped from the cluster, 
the integral is restricted to the domain $p^2/2m_\nu+m_\nu \varphi(r)<m_\nu \varphi(\infty)\equiv 0$.
The gravitational potential $\varphi$ is solved from
 the Poisson equation $\varphi''+2\varphi'/r=4\pi G\rho$, with the total mass density $\rho=\rho_G+\rho_g+\rho_\nu$.

The data in fig. \figSig \, expose that $\Sigma$ does not decay fast for $r>300$ kpc.
Hence our previous fits  \cite{nieuwenhuizen2009non,nieuwenhuizen2011prediction,nieuwenhuizen2013observations}
with a large chemical potential $\mu$ and  fast decay of $\rho_\nu(r)$ do not apply.
But this turns out to be a blessing: isothermal fermions  can produce a good fit to the data  {\it including the tails}.
Then, when plotting the baryon fraction 
$f_b=M_b/(M_\nu+M_b)$, with $M_b=M_G+M_g$, as function of $r$, 
we notice around $\rms= 2250$ kpc a quadratic maximum
$f_b(\rms)=0.16\pm0.01$, overlapping with the cosmic value $f_b^c=0.1580\pm0.0014$ \cite{planck2015planck}. 
The absence of a ``missing baryon problem'' around $\rms $ allows to eliminate it theoreticallly:
we take the gas density from the observations and continue its fit (\ref{nefit}) beyond 1 Mpc,
solve $\rho_\nu$ from the Poisson equation up to some {\it mass separation radius} $\rms $ and
impose beyond $\rms $  the cosmic ratio $\rho_\nu(r)=(1/f_b^c-1) \rho_g(r)$,
so that it holds there that $M_\nu(r)=(1/f_b^c-1) M_b(r)$  and $f_b(r)=f_b^c$ (see Fig. 3).
To achieve the matching, we set $f_b^c=\bar f_b^c\pm\delta \! f_b^c$  and add to $\chi^2(\Sigma,g_t,n_e)$ the combination

\BEQ\label{point31+32}
\chi_{\it ms}^2(r_\ms)=
\frac{1}{\delta\! f_{b}^{c}{}^2}
\left(\frac{M_b}{M}-\bar f_{b}^c\right)^2+
\frac{1}{\delta\! f_{b}^{c}{}^2}\left(\frac{\rho_b}{\rho}-\bar f_{ b}^c\right)^2,
\EEQ
with  $M_b=\fgb M_G+M_g$, $M=M_\nu+M_b$, all taken at $r_{\it ms}$, and likewise for the $\rho$'s.
$\fgb$ is the baryon fraction of the BCG, which we take as $1$, though $f_b^G=\bar f_b^c$ would hardly change the fit.
The best minimum yields $\chi^2(\Sigma,g_t,n_e)+\chi^2_{\it ms}(r_\ms)=131.9$ for 
$\nu=97-11$ d.o.f., so that its $q=0.0011$ slightly improves the gas-only value.
The  $\chi^2_\sl(\Sigma)=5.63$, $\chi^2_\wl(\Sigma)=8.33$, $\chi^2_\wl(g_t)=28.8$
add up to $\chi^2(\Sigma,g_t)= 42.7$. Their $\nu=39-6$ d.o.f. imply
 $\chi^2(\Sigma,g_t)/\nu =1.29$ and the good lensing-alone $q$-value $0.12$.
The fits  for $\Sigma$ and $g_t$ are represented in figs. \figSig \,and \figgt, respectively.

Accounting for the variation of  $r_\ms$ causes conservative error bars.
The BCG has mass and inner and outer radius

\BEQ    
&& 
\hspace{-7mm} M_G=  3.2\pm 1.0 \,10^{13}M_\odot, \qquad 
R_G^{\, i}  =7.2  \pm 1.4\, \kpc,  \qquad  R_G^o = 129\pm 40\,\kpc. 
\EEQ   
The gas fit (\ref{gasfit}) does not get altered within the error bars when combining its data with the lensing data.
The velocity dispersion and chemical potential read

\BEQ 
\sigma_\nu=1330\pm 40\,\frac{{\km}}{{\s}}, \,\,\,
\mu-\varphi(0)=5.0\pm 0.9 \, \,10^6\frac{\km^2}{\s^2}.
\EEQ
Finally, the from (\ref{nudens}) the fermion mass comes out in the combination $gm_\nu/\lambda_\nu^3\sim gm_\nu^4$ 
due to the thermal length $\lambda_\nu =\sqrt{2\pi}\hbar/m_\nu\sigma_\nu$. The fit yields

\BEQ  \label{m-nu-g}
\left(\frac{\gbar}{12}\right)^{1/4}m_\nu =1.92_{-0.16}^{+0.13} \,  \frac{{\rm eV}}{c^2}.
\EEQ
This exceeds the $1.45\pm0.03$ of \cite{nieuwenhuizen2009non}, the $1.55\pm0.04$ of  \cite{nieuwenhuizen2011prediction}
and the $1.51\pm 0.04$ \eVct \  of \cite{nieuwenhuizen2013observations}.
However, in those works an additional type of dark matter is assumed, and modeled explicitly in Ref. \cite{nieuwenhuizen2013observations}.

The depth of the potential well $\varphi(0)=-19.4\pm0.4 \,\,10^6$ km$^2/$s$^2$
has also a comparable systematic error from the extrapolation of the gas data beyond 1 Mpc  with the S\'ersic profile. 

At $r_{200}=2.37$ Mpc the cluster has overdensity $200$ with 
$M_{200}=(18.1 \pm 0.8)10^{14}M_\odot$, to be
compared with the $(1.32\pm 0.09)/h\,10^{15}M_\odot=(18.9\pm 1.3)$ $10^{14}M_\odot$ of \cite{umetsu2015three}.

At $\rms=2.05\pm0.39$ Mpc the Newton acceleration 
\BEQ
\frac{GM(\rms)}{\rms^2}=5.4\,10^{-11}\frac{{\rm m}}{{\rm s}^2}=0.92\frac{cH_0(\bar z)}{4\pi},
\EEQ
with $M(\rms)=(16.4 \pm 0.7)10^{14}M_\odot$,
leads to a consistent picture: Beyond $\rms $ the cosmic acceleration dominates the attraction to the cluster, preventing mass  separation.
Hence, unlike $r_{200}$,  $\rms$ has a clear physical meaning.

\begin{figure}
\label{figSLdataLiUm}
\centerline{\includegraphics[scale=1]{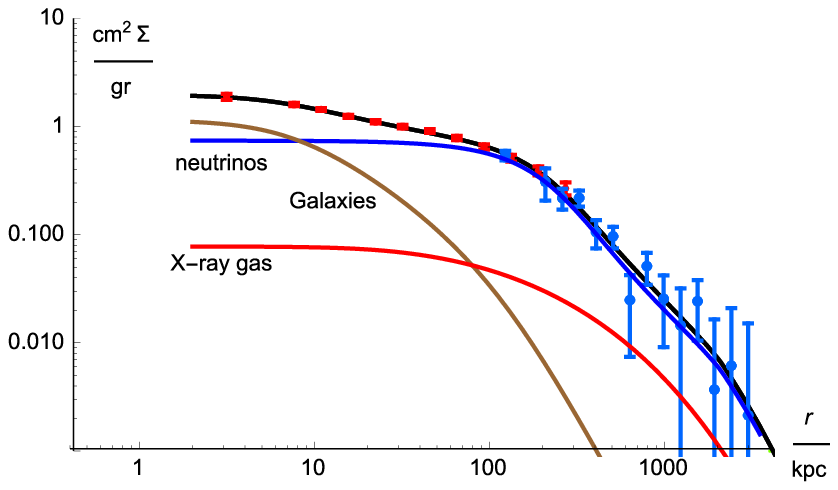} \vspace{-12mm} }
\centerline{ \includegraphics[scale=1]{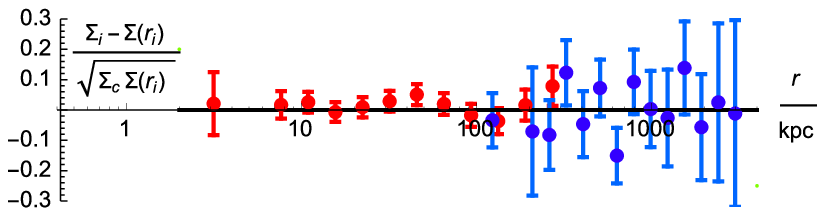}}
 \caption{$2d$ mass density $\Sigma$ and the contributions from its components as function of $r$.
Left data from \cite{limousin2007combining} with error bars $(\overline C^\sl_{ii})^{1/2}$. 
Right data from  \cite{umetsu2015three}  with error bars $(C^\wl_{ii})^{1/2}$.
Black line: best fit to our model. The kink in the slope at $r_\ms=2.05$ Mpc reflects the mass separation: below $r_\ms$
the neutrinos deplete towards the interior. Lower pane: Relative deviations.}
\end{figure}

\begin{figure}
\label{figWLUBdataa}
\centerline{ \includegraphics[scale=1]{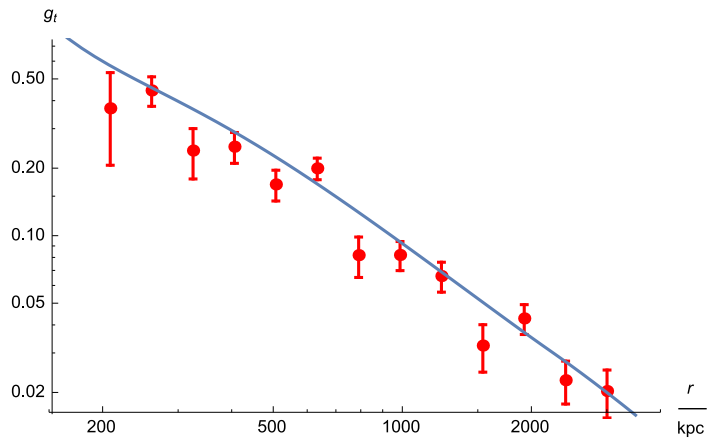}}
 \caption{
Weak lensing shear $g_t$ as function of $r$.
Data from Umetsu et al. \cite{umetsu2015three}; full line: best fit to the model.}
\end{figure}

\begin{figure}\label{figfB}
\centerline{\includegraphics[scale=1] {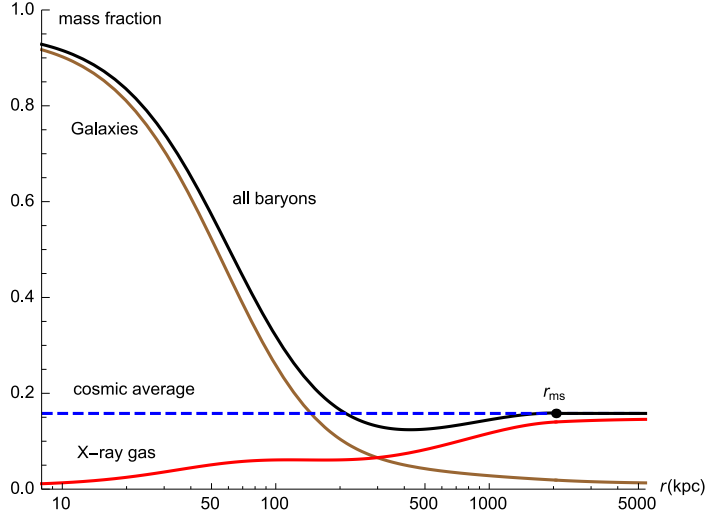}}
\caption{Baryonic mass fractions as function of $r$. In the centre the brightest cluster galaxy dominates.
The depletion beyond 300 kpc is standard.
Beyond 700 kpc the neutrinos are relatively depleted, allowing the baryons to reach the cosmic fraction 0.158
at the mass separation radius $\rms=2.05$ Mpc. From there on baryons and neutrinos occur at cosmic ratios.}
\end{figure}

\begin{figure}\label{pTdata}
\centerline{ \includegraphics[scale=1] {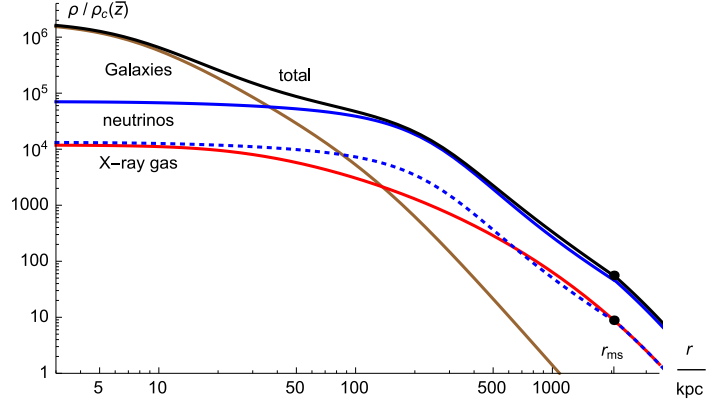}}
\caption{Mass densities of Galaxies, X-ray gas and neutrinos in A1689, normalised to the critical density $\rho_c(\bar z)$, as function of $r$.
The dotted line presents the gas density if it had the cosmic fraction of the neutrino mass density.
Its overlap beyond $r_\ms$ indicates that our model avoids a missing baryon problem; its near-overlap at the origin a is surprise.
The surplus from 20 to   650 kpc and the depletion from 650 to $\rms=2050$ kpc show that the neutrinos are  cored.}
\end{figure}

\begin{figure}\label{figfB}
\centerline{\includegraphics[scale=1] {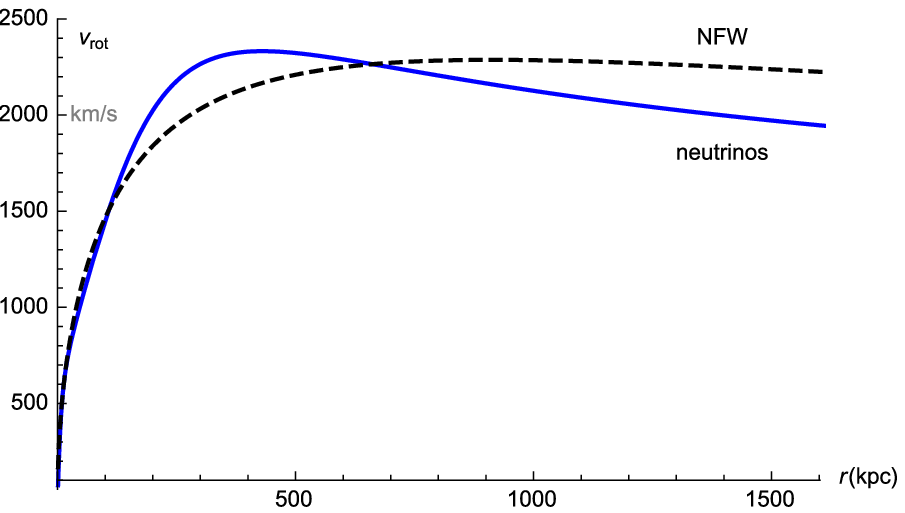}}
\caption{
Full line: Circular rotation velocity of galaxies as function of $r$; 
the bump around 430 kpc is caused by the neutrino core. Dashed: NFW prediction. }
\end{figure}

\subsection{The cosmic budget and neutrinoless double beta decay}
The mass (\ref{m-nu-g}) depends on the degeneracy factor $\gbar $:
In the cluster a smaller DM particle mass can be compensated by a larger degeneracy, at fixed $\gbar ^{1/4}m$.
At the cosmic scale $\gbar m$ gets fixed, lifting the degeneracy.
Each active (anti)neutrino has a number density 
\masscorr{{\bf Moppie: of  56.0 $\mapsto$}} 
$n_\nu^a=55.86$  $\cm^{-3}$ and we assume that sterile $\nu$'s are also thermal.
Supposing there are $N_f=3+\Delta N_f$ neutrino families with $\gbar =2N_f$ degrees of freedom, all having basically a mass of 1.92 eV, 
the cosmic budget comes out as 
$\Om_\nu=({N_f}/{6})^{3/4}(0.251\pm0.019)h_{70}^{-3/2}$,
where we reinserted Hubble's constant  \cite{nieuwenhuizen2009non}. 
The physical density parameter thus reads
$ \Om_\nu h^2=({N_f}/{6})^{3/4}(0.123 \pm 0.009)h_{70}^{1/2}$.
The Planck cold dark matter fraction is $\Om_ch^2=0.1188\pm0.0010$ \cite{planck2015planck}.
Neutrinos can exactly account for this in case $ N_f=6$, $\gbar =12$, which amounts to the 3 active and 3 sterile ones 
(3+3 model), all having nearly  the same mass. 
That case for $\Delta N_f=2$ needs $m_\nu=2.14$ \eVct \, which is ruled out \cite{Agashe:2014kda}.

In general it is not required that the masses are equal, and neither the chemical potentials nor the velocity dispersions.
While neutrino oscillations have shown small mass differences between the active neutrinos, reactor experiments 
expose hints for one, two or possibly 3 sterile neutrinos with eV mass differences;
at least two sterile species are needed to explain CP violation \cite{giunti2007fundamentals,schwetz2011global,kopp2011there}. 
Neutrinoless double beta decay ($0\nu\beta\beta$) implies for the known mixing angles \cite{Agashe:2014kda} a sharpest lower bound 
when both Majorana phases are equal to $\pi$, namely $m_{\beta\beta}^{0\nu}\ge 0.33m_\nu=0.64\pm0.05$ \eVct.
Upper bounds for $m_{\beta\beta}^{0\nu}$ are $0.45\,\eV$ by EXO-200, $0.28\,\eV$ 
by KamLAND-Zen \cite{asakura2015results}, $0.4\,\eV$ by GERDA \cite{majorovits2015search} and 
of $0.27-0.76$ eV by CUORE \cite{Cuore2015}, respectively. 
  Hence Majorana neutrinos are disfavoured and we arrive at Dirac neutrinos:  
uncharged electrons with different flavour and mass eigenbases, like quarks (or: quasi-Dirac neutrino's, nearly having these properties.)
Detection of $0\nu\beta\beta$, though, would lead to a smaller mass of the active $\nu$'s and a higher mass(es) of the sterile $\nu$'s.

We considered only neutrinos bound to the cluster.  Accounting, as in our earlier works,  also for the unbound ones would 
overlook that they probably had time enough to escape to infinite.
The case yields anyhow a mass and other parameters compatible within the error bounds.

\masscorr{ {We noticed the overlap of our $\Omega_\nu$ with the $\Omega_c$ from the CMB. Equating them and taking the smaller error bar of the latter
\cite{planck2015planck} offers a sharper prediction for the 3+3 Dirac mass: }}

We noticed the overlap of our 
\masscorr{ {\bf Replace $\Omega_\nu$ by:}}
 $\Omega_\nu=gm_\nu n_\nu^a/\rho_c$ with the $\Omega_c$ from the CMB. Equating them and taking the smaller error of the latter
\cite{planck2015planck} offers a sharper prediction for the 3+3 Dirac \masscorr{ {\bf replace: ``mass:''  by}} 
mass. The effective number of active neutrino families $N_{\! f}^a=3.046$ \cite{cyburt2015big} yields 
$g=12(N_{\! f}^a/3)^{3/4}=12.14$, implying a sharp prediction for their mass,

\BEQ 
\masscorr{ m_\nu=(1.861\pm0.016)\frac{1}{h_{70}^2}\frac{\eV}{c^2}.     \mapsto  \textrm{{\bf THIS goes to }}}
 m_\nu=(1.847\pm0.016)\frac{\eV}{c^2}, \EEQ 
 \masscorr{{\bf I ADD: which is independent of the Hubble parameter $H_0$.}}
which is independent of the Hubble parameter $H_0$.
The Dirac nature implies that the $\Delta m^2\sim 1 $ \eVct$^2$ \, effects seen so far at 2$\sigma$--3$\sigma$  
evidence in neutrino oscillations  must  be statistical flukes. 
Much effort  will be needed to test this prediction and establish  its Dirac mass differences
$\Delta m_{12}=1.1\,10^{-5}m_\nu$ and $|\Delta m_{13}|=3.5\,10^{-4}m_\nu=1.3\,10^{-9}m_e$. 
 
It is generally expected that active neutrinos have sub-eV masses,
implying that they hardly contribute to the cosmic mass budget.
Within our approach this case leads to 5 or 6 sterile neutrino families (3+5, 3+6 models).
These extra modes are massless during big bang nucleosynthesis, so while they may
put stress on observations of $^4$He and D, they may actually help to soften the $^7$Li problem \cite{cyburt2015big}.

Massive neutrinos are known not to describe the CMB correctly. Our 3+3 Dirac model with 
\masscorr{$m_\nu =1.86\,/h_{70}^{2}$ eV {\bf THUS modifies towards }}
$m_\nu =1.85$ eV, can be fit  to the Planck CMB data \cite{planck2015planck}.
With the CLASS-code \cite{lesgourgues2011cosmic} we find that the parameter set  
$\Omega_bh^2 = 0.0245$, $h=0.725$, 
$\tau_{\rm reio} = 0.175$, $A_s = 2.19\,10^{-9}$, $n_s = 1.33$ and  $Y_{\rm He} = 0.25$
 achieves to have the peaks at the right positions with amplitude at  70\% of the first acoustic TT peak and 
within some relative 10\% below or above the other TT, TE and EE peaks.
It remains to be seen whether the deviations can be repaired by the effects of turbulence.

\section{Summary}

The dark matter of the galaxy cluster A1689 is modelled by isothermal neutrinos. 
The fit works  well and is remarkably consistent,  without ``missing baryons''.
Beyond the {\it mass separation radius} $\rms =2.1$ Mpc the baryon fraction is cosmic, consistent observations \cite{morandi2015galaxy};
 the Newton force, weaker than  the cosmic acceleration $cH_0/4\pi$, is unable to achieve mass separation.
When for $r<r_\ms$ the actual ratio of the mass densities of the $\nu$'s and the X-ray gas is compared to the cosmic ratio, 
the ``excess mass'' of the $\nu$-core is compensated in the outskirts \cite{morandi2015galaxy} by
an $\nu$  under-concentration between 0.7 and 2.1 Mpc, rather than by a baryon over-con\-centration \cite{rasheed2011searching}.

The rotation around the cluster centre  has a maximum of 2330 km/s at $r=430$ kpc due to the neutrino core (Fig. 6), 
a further test for the model. The central X-ray gas density appears to be near the cosmic baryon ratio w.r.t. the local neutrino 
dark matter density (Fig. 5).

It is desirable to have gas data beyond 1 Mpc, so that the extrapolation employed in our model is better constrained. 
Likewise, a new strong lensing analysis may update the 2007 Limousin et al result \cite{limousin2007combining},
which is central for sharply fixing the model parameters,  and known to lead to NFW fits different from the ones for weak lensing.

Our cluster results stem with the cosmic ones if there exist 3 Dirac neutrino families with mass of approximately 
\masscorr{ 1.86 eV. {\bf NoNo}} 
1.85 eV/$c^2$. This is not automatically ruled out by free streaming because the plasma may have been turbulent, as 
expected from its high Reynolds number and from the stretched exponential shape
of the galaxy power spectrum \cite{bershadskii2015deterministic}. 
Perhaps the  $^7$Li problem \cite{cyburt2015big} can be solved by small-scale temperature differences arising from turbulence.
Support for the neutrino picture may come from the ``cosmic train wreck" galaxy cluster Abell 520,
where a central clump with huge mass-to-light ratio $800M_\odot/L_{R\odot}$ exhibits DM  separated from the galaxies
\cite{jee2012study,clowe2012dark,jee2014hubble}. This is a puzzle for $\Lambda$CDM, but colliding
cores of degenerate neutrinos may partly end up in the center due to the exclusion principle.
After all, ``neutrino stars'' (i.e., neutrino cores) may act as nuclei in heavy ion collisions.
Likewise, such cores may hinder each other and cause an offset between baryonic and dark mass as 
in Abell 3827 \cite{massey2015behaviour}, where it is larger than \LCDM \  can explain \cite{schaller2015offsets}.

Neutrinos of mass 
\masscorr{ 1.86 eV. {\bf NoNo}} 
1.85 eV$/c^2$ become nonrelativistic at redshift 7900, so near the recombination they 
behave like cold dark matter and explain most of the CMB data, up to some 10\% accuracy.
However, the standard theory of linear fluctuations does not account for turbulence effects,
so that a match may still arise. The KATRIN experiment searches the neutrino mass down to 0.2 eV/$c^2$. 
Most likely they are Dirac particles with mass in the $\bar \nu_e$ itself, so that the detection or rejection of our 
prediction will be achievable.

\vspace{5mm}
{\it Acknowledgments:}
We thank Marceau Limousin, Andrea Morandi and Keiichi Umetsu  for supplying data,
and  Keiichi Umetsu, Andrea Morandi, 
Armen Allahverdyan, Erik van Heusden and Remo Ruffini for discussion. 

\vspace{5mm}

\bibliographystyle{iopart-num} 

\bibliography{references-A1689-EmQM}

\end{document}